\documentclass[11pt]{article}
\usepackage{mathrsfs}
\usepackage{amssymb}
\usepackage{color}
\usepackage{amsmath}
\usepackage{ascmac}
\usepackage{amsthm}
\usepackage[dvips]{graphicx}
\usepackage{booktabs}

\newtheorem{thm}{Theorem}

\def\be{{\beta}}
\def\ga{{\gamma}}

\def\la{{\lambda}}
\def\si{{\sigma}}

\def\th{{\theta}}

\def\bbe{{\text{\boldmath $\beta$}}}

\def\bta{{\text{\boldmath $\eta$}}}

\def\bsi{{\text{\boldmath $\sigma$}}}

\def\bth{{\text{\boldmath $\theta$}}}

\def\bphi{{\text{\boldmath $\phi$}}}

\def\bmu{{\text{\boldmath $\mu$}}}

\def\thh{{\widehat \th}}

\def\sih{{\widehat \si}}

\def\0{{\text{\boldmath $0$}}}

\def\k{{\text{\boldmath $k$}}}

\def\w{{\text{\boldmath $w$}}}

\def\z{{\text{\boldmath $z$}}}
\def\A{{\text{\boldmath $A$}}}

\def\E{{\text{\boldmath $E$}}}

\def\H{{\text{\boldmath $H$}}}
\def\I{{\text{\boldmath $I$}}}

\def\P{{\text{\boldmath $P$}}}

\def\R{{\text{\boldmath $R$}}}

\def\X{{\text{\boldmath $X$}}}

\def\Z{{\text{\boldmath $Z$}}}

\def\1{{\text{\boldmath $1$}}}

\def\Re{\mathbb{R}}
\def\Var{{\rm Var}}

\def\E{{\rm E}}
\def\sgn{{\rm sgn}}

\def\Ga{\Gamma}

\title{Bayesian Estimators for Small Area Models Shrinking Both Means and Variances}

\author{
Shonosuke Sugasawa\thanks{The Institute of Statistical Mathematics}, 
\ Hiromasa Tamae\thanks{Graduate School of Economics, University of Tokyo}
\ and Tatsuya Kubokawa\thanks{Faculty of Economics, University of Tokyo}
}

\date{}

\begin{document}

\maketitle

\begin{abstract}
For small area estimation of area-level data, the Fay-Herriot model is extensively used as a model based method.
In the Fay-Herriot model, it is conventionally assumed that the sampling variances are known whereas estimators of sampling variances are used in practice.
Thus, the settings of knowing sampling variances are unrealistic and several methods are proposed to overcome this problem.
In this paper, we assume the situation where the direct estimators of the sampling variances are available as well as the sample means.
Using these information, we propose a Bayesian yet objective method producing shrinkage estimation of both means and variances  in the Fay-Herriot model.
We consider the hierarchical structure for the sampling variances and we set uniform prior on model parameters to keep objectivity of the proposed model.
For validity of the posterior inference, we show under mild conditions that the posterior distribution is proper and has finite variances.
We investigate the numerical performance through simulation and empirical studies.

\par\vspace{4mm}
 \end{abstract}

\section{Introduction}
Small area estimation has been a topic of great interest to applied and theoretical statisticians in recent years. 
The reliability of small area estimates is an essential issue for making useful policy decisions. 
It is well known that the direct survey estimates for small areas are usually unreliable, having large standard errors and coefficients of variation. 
Therefore, it is necessary to use statistical models to connect the related small areas, and obtain estimates with improved precision by `borrowing strength' across areas. 
For reviews over the techniques of small area estimation, we refer to Ghosh and Rao (1994), Pfeffermann (2002) and Rao and Molina (2015).

A famous small area model for treating area-level data is the Fay-Herriot model suggested by Fay and Herriot (1979).
In the Fay-Herriot model, it is conventionally assumed that the sampling variances are known.
In practice, however, the sampling variances are often estimated in various ways, and the small area estimators are provided by replacing the known variances with their estimators.
This means that the small area estimators derived in the Fay-Herriot model involve substantial errors which come from estimation of variance, and we need to evaluate the estimation errors.
To this end, several approaches are developed in the small area literature.
We refer to Arora and Lahiri (1997), You and Chapman (2006), Liu, Lahiri and Kalton (2007), Wang and Fuller (2003), Rivest and Vandal (2003), Otto and Bell (1995), Huff, Eltinge and Gershunskaya (2002), Cho, Eltinge, Gershunskaya, Huff (2002) and Eltinge, Cho and Hinrichs (2002).
In these papers, much attention has been paid to accounting sampling variance estimation effectively.

You and Chapman (2006) proposed the modified Fay-Herriot model taking the estimated sampling variance into the Fay-Herriot model.
To describe their model, suppose that there are $m$ small areas, and let $(X_i,S_i^2)$ be a pair of direct survey estimates of mean and variance in the $i$-th small area for $i=1,\ldots,m$.
Let $\z_i=(z_{i1},\ldots,z_{ip})'$ be a vector of $p$ covariates available at the estimation stage. 
Then the Fay-Herriot model can be modified as
\begin{equation}\label{YC}
\begin{split}
&X_i|\theta_i,\si_i^2\sim N(\theta_i,\si_i^2), \ \ \ \ \theta_i\sim N(\z_i'\bbe,\tau^2)\\
&S_i^2|\si_i^2\sim \Ga\left(\frac{n_i-1}{2},\frac{n_i-1}{2\si_i^2}\right), \ \ \ \ \si_i^2\sim \pi(\si_i^2)
\end{split}
\end{equation}
where $(X_i, S_i^2, \theta_i, \si_i^2)$, $i=1, \ldots, m$, are mutually independent and $\Ga(a,b)$ denotes the gamma distribution with density proportional to $x^{\alpha-1}\exp(-\beta x), \ x>0$.
Here, $n_i$ is the sample size for a simple random sample in the $i$-th area, $\bbe=(\beta_1,\ldots,\beta_p)'$ is the $p\times 1$ vector of regression coefficients.
In the framework of (\ref{YC}), You and Chapman (2006) suggested the hierarchical Bayesian approach by setting prior distributions:
$$
\pi(\bbe)\propto 1, \ \ \si_i^{2}\sim IG(a_i,b_i), \ i=1,\ldots,m, \ \ \ \ \tau^{2}\sim IG(a_0,b_0),
$$  
where $IG(a,b)$ is the inverse Gamma density function with density proportional to $x^{-\alpha-1}\exp(-\beta/x)$, $x>0$, and $a_i,b_i \ (i=0,\ldots,m)$ are chosen to be very small known constants, so that the prior distributions on $\si_i^2$ and $\tau^2$ are close to the uniform distribution.
However, the nearly uniform prior distribution for $\si_i^2$ does not produce shrinkage estimation of the sampling variances.

On the other hand, recently, Maiti, Ren and Sinha (2014) proposed the empirical Bayes approach for (\ref{YC}), namely
\begin{equation}\label{maiti}
\begin{split}
&X_i|\theta_i,\si_i^2\sim N(\theta_i,\si_i^2), \ \ \ \ \theta_i\sim N(\z_i'\bbe,\tau^2)\\
&S_i^2|\si_i^2\sim \Ga\left(\frac{n_i-1}{2},\frac{n_i-1}{2\si_i^2}\right), \ \ \ \ \si_i^2\sim IG(\alpha,\gamma),
\end{split}
\end{equation}
where $\bbe,\tau^2,\alpha$ and $\gamma$ are unknown parameters.
They estimated model parameters $\bbe$ and $\tau^2$ as well as $\alpha$ and $\gamma$ from the (marginal) likelihood function.
However, the marginal likelihood function cannot be obtained in a closed form and they developed the EM algorithm for getting estimates of the model parameters.
Also we found through the simulation study that the estimates of $(\gamma,\alpha)$ tend to be unstable.
Moreover, the analytical expression of the Bayes estimator of $\theta_i$ is hard to obtain since the posterior distribution of $\theta_i$ is no longer a normal distribution but an unfamiliar distribution.
Thus, it is worth developing much easier yet practical method shrinking both means and variances in small area estimation.

These observations motivate us to propose the Bayesian approach for small area models shrinking both mean and variances.
To achieve this, we assume the uniform prior distributions on $\tau^2$ and $\bbe$, namely $\pi(\bbe,\tau^2)\propto 1$, and the following structure is introduced for $\si_i^2$:
$$
\si_i^2\sim IG(a_i,b_i\gamma), \ \ \ i=1,\ldots,m, \ \ \ \ \pi(\gamma)\propto 1,
$$
where $a_i$ and $b_i$ are constants specified by users.
A suggestion for the choice of $a_i$ and $b_i$ is given in the end of Section \ref{sec:B1}.
In these settings, the full conditional posterior distributions are all familiar forms that enable us to easily draw the samples via the Markov chain Monte Carlo technique, in particular the Gibbs sampler as discussed in Section \ref{sec:Bayes}.
Using these posterior samples, we obtain the point estimates of the parameter of interest $\theta_i$ by the simple average of posterior samples.
Moreover, the prediction intervals are easily constructed from quantiles of posterior samples compared to the empirical Bayes confidence intervals given in Dass, Maiti, Ren and Sinha (2012) and Hwang, Qiu and Zhao (2009).
In Section \ref{sec:var}, we also consider the alternative formulation of the true variance $\si_i^2$ in each area with use of covariate information, namely $\si_i^2$ is structured as $\si_i^2\sim IG(a_i,b_i\gamma\exp(\w_i'\bta))$ for some vector of covariates $\w_i$ and unknown regression vector of coefficients $\bta$.
In this paper, we also develop a Bayesian method for this model and prove the posterior propriety and finiteness of the posterior variances when we use the improper priors for unknown parameters.

The paper is organized as follows: In Section \ref{sec:Bayes}, the full Bayesian model alternative to Maiti, et al. (2014) and You and Chapman (2006) is proposed. 
The full conditional distribution is described, and the Gibbs sampling for MCMC is given.
As a theoretical main result, under a mild sufficient condition, we prove that the resulting posterior distribution is proper and the model parameters have finite variances.
In Section \ref{sec:num}, we carry out simulation studies to compare the suggested methods with the models by Maiti, et al. (2014) and You and Chapman (2006).
As real data analysis, we apply our methods to two real data sets, the SFIE data in Japan and the famous corn crop data.
The concluding remarks are given in Section \ref{sec:conc} and the proofs are given in the Appendix.

\section{Bayesian models shrinking both means and variances}\label{sec:Bayes}

\subsection{Model settings and Bayesian inferences}\label{sec:B1}
We propose Bayesian multi-stage small area model shrinking both means and variances described as
\begin{equation}\label{model}
\begin{split}
&X_i|\theta_i,\si_i^2\sim N(\theta_i,\si_i^2), \ \ \ \ \theta_i|\bbe,\tau^2\sim N(\z_i'\bbe,\tau^2), \\
&S_i^2|\si_i^2\sim \Ga\left(\frac{n_i-1}{2},\frac{n_i-1}{2\si_i^2}\right), \ \ \ \si_i^2|\ga\sim IG(a_{i},b_{i}\ga)\\
&\pi(\bbe,\tau^2,\ga)=1,
\end{split}
\end{equation}
where $(X_i, S_i^2, \theta_i, \si_i^2)$, $i=1, \ldots, m$, are conditionally independent given $(\bbe, \tau^2, \ga)$.
Here, $a_i,b_i$ are positive and known (user specified) constants.
The choice of $a_i$ and $b_i$ is not concerned with the propriety of the posterior distributions given in Theorem
\ref{thm:pos} as far as $a_i$ and $b_i$ are positive.
The practical choice of these constants is discussed later.
Note that the model for $S_i^2$ in (\ref{model}) means that $(n_i-1)S_i^2/\si_i^2$ given $\si_i^2$ follows a chi-square distribution with $(n_i-1)$ degrees of freedom.
This setting is appropriate under simple random sampling, but for complex sampling design, the degrees of freedom needs to be determined carefully as discussed in Maples, Bell and Huang (2009).

We now consider the posterior distribution and investigate its properties. 
We denote $D=\{X_i,S_i^2,\z_i\}_{i=1,\ldots,m}$, the set of all observed data, for notational simplicity.
From the formulation (\ref{model}), the posterior density is given by
\begin{equation}\label{pos}
\begin{split}
&\pi(\theta_1,\ldots,\theta_m,\si_1^2,\ldots,\si_m^2,\bbe,\tau^2,\ga|D)\\
&\ \ \propto (\tau^2)^{-m/2}\prod_{i=1}^{m}\ga^{a_i}(\si_i^2)^{-n_i/2-a_i-1}\exp\left\{-\frac{(X_i-\theta_i)^2+(n_i-1)S_i^2+2b_i\ga}{2\si_i^2}-\frac{(\theta_i-\z_i'\bbe)^2}{2\tau^2}\right\}.
\end{split}
\end{equation}
We state our main result, which provides a sufficient condition for the propriety of the posterior distribution.
To this end, we define $\Z=(\z_1,\ldots,\z_m)$.

\begin{thm}\label{thm:pos} 
{\rm  (a)}\ \ The marginal posterior density $\pi(\bbe,\tau^2,\ga|D)$ is proper if $m>p+2$, $n_i>1$ and ${\rm rank}(\Z)=p$.\\
{\rm (b)}\ \ The model parameters $\bbe,\tau^2$ and $\ga$ have finite posterior variances if $m>p+6$, $n_i>1$ and ${\rm rank}(\Z)=p$.
\end{thm}

Part (a) of Theorem \ref{thm:pos} says that the marginal posterior densities of the small area means are proper and part (b) establishes a sufficient condition for obtaining finite measures of uncertainty for the model parameters. 
We note that the sufficient condition given in Theorem \ref{thm:pos} is the same as the condition given in Arima, Datta and Liseo (2015) except for $n_i>1$, where they suggested Bayesian estimators for small area models with measurement errors in covariates.
The proof of Theorem \ref{thm:pos} is deferred to the Appendix.

Since the posterior distribution in (\ref{pos}) cannot be obtained in a closed form, we rely on the Markov chain Monte Carlo technique, in particular the Gibbs sampler, in order to draw samples from the posterior distribution.
This requires generating samples from the full conditional distributions of each of $(\theta_1,\ldots,\theta_m,\si_1^2,\ldots,\si_m^2,\bbe,\tau^2)$ given the remaining parameters and the data $D$. 
From the expression given in (\ref{pos}), the full conditional distributions are given by
\begin{equation}\label{full}
\begin{split}
&\theta_i|\bbe,\tau^2, \bsi^2, \bth_{(-i)}, \ga, D\sim N\left(\frac{\tau^2 X_i+\si_i^2\z_i'\bbe}{\tau^2+\si_i^2},\frac{\tau^2\si_i^2}{\tau^2+\si_i^2}\right), \ \ \ i=1,\ldots,m\\
&\si_i^2|\bbe,\tau^2, \bsi_{(-i)}^2, \bth,\ga,D\sim IG\left(\frac{n_i}{2}+a_i,\frac12(X_i-\theta_i)^2+\frac12(n_i-1)S_i^2+b_i\ga \right), \ \ \ i=1,\ldots,m\\
&\bbe|\tau^2, \bsi^2, \bth,\ga,D\sim N_p\left((\Z'\Z)^{-1}\Z'\bth,\tau^2(\Z'\Z)^{-1}\right),\\
&\tau^2|\bbe,\bsi^2, \bth,\ga,D\sim IG\left(\frac{m}{2}-1,\frac12(\bth-\Z\bbe)'(\bth-\Z\bbe)\right),\\
&\ga|\bbe,\tau^2,\bsi^2, \bth,D\sim \Gamma\left(\sum_{i=1}^ma_i+1,\sum_{i=1}^m\frac{b_i}{\si_i^2}\right),
\end{split}\end{equation}
where $\bsi^2=(\si_1^2, \ldots, \si_m^2)'$, $\bth=(\theta_1,\ldots,\theta_m)'$, and the suffix $(-i)$ denotes the vector without the $i$-th component.
Fortunately, the full conditional distributions for every parameter are familiar distributions allowing us to easily implement the Gibbs sampling.

In closing of this section, we give a suggestion for the choice of $a_i$ and $b_i$. 
For fixed value of $\gamma$, it is noted that
\begin{align*}
\Var(X_i)=\E[\Var(X_i|\th_i)]+\Var(\E[X_i|\th_i])=\E[\si_i^2]=\frac{b_i}{a_i-1}\gamma.
\end{align*}
Since $X_i$ is the sample mean, it is natural to consider $\Var(X_i)=O(n_i^{-1})$. 
On the other hand, the full conditional expectation of $\si_i^2$ is obtained from (\ref{full}) as
\begin{align*}
E[\si_i^2|X_i,\th_i,S_i^2]
&=\frac{(X_i-\th_i)^2/2+(n_i-1)S_i^2/2+b_i\gamma}{n_i/2+a_i-1}\\
&=\frac{n_i/2}{n_i/2+a_i-1}{\widetilde\sigma}^2_i(X_i,S_i^2)+\frac{a_i-1}{n_i/2+a_i-1}\cdot \frac{b_i}{a_i-1}\gamma
\end{align*}
where
$$
{\widetilde\sigma}^2_i(X_i,S_i^2)=\frac{1}{n_i}\left\{(X_i-\th_i)^2+(n_i-1)S_i^2\right\}.
$$
Thus the full conditional expectation of $\si_i^2$ is the weighted mean of ${\widetilde\sigma}^2_i(X_i,S_i^2)$ and the prior mean $b_i\gamma/(a_i-1)$, and the weight for the prior mean is determined by $a_i$.
It is natural that the posterior full conditional expectation approaches to $S_i$ for large $n_i$.
Thus it is reasonable to choose $a_i$ as $a_i=O(1)$ for $n_i$. 
These observations show that the order of $a_i$ and $b_i$ should be $a_i=O(1)$ and $b_i=(n_i^{-1})$.
Hence, we suggest to use $a_i=2$ and $b_i=n_i^{-1}$ as the one reasonable choice.
In the simulation and empirical studies given in the subsequent section, we use these values for $a_i$ and $b_i$.
In empirical study, we investigate the influence of choices of $a_i$ and $b_i$.

\subsection{Alternative formulation of heteroscedastic variances}\label{sec:var}
We next suggest the alternative formulation of heteroscedastic variances $\si_i^2$ in each area.
Remember that we assume that $\si^2_i\sim IG(a_i,b_i\ga)$ for specified $a_i$ and $b_i$ in the previous subsection.
However, in case that we can accommodate the covariate information in the variance modeling, more sophisticated modeling can be developed. 
Let $\w_i$ be a vector of $q$ covariates in the $i$-th area and $\bta$ is a $q$-dimensional vector of unknown coefficients, and we propose the structure $\si_i^2\sim IG\left(a_i,b_i\ga\exp(\w_i'\bta)\right)$ with typical choice $a_i=2$ and $b_i=1/n_i$.
Let $\w_i=(w_{i1},\ldots,w_{iq})'$ and $\bta=(\eta_1,\ldots,\eta_q)'$, then we cannot assign $w_{i1}=1$ for $i=1,\ldots,m$ since we cannot identify $\ga$ and $\eta_1$ in this case.
To develop a Bayesian inference, we again use the uniform prior distribution for all parameters $\bbe,\tau^2,\ga$ and $\bta$, namely $\pi(\bbe,\tau^2,\ga,\bta)\propto 1$, to keep objectivity of inferences.
Therefore, the covariate dependent version of (\ref{model}) is given by
\begin{equation}\label{var2}
\begin{split}
&X_i|\theta_i,\si_i^2\sim N(\theta_i,\si_i^2), \ \ \ \ \theta_i|\bbe,\tau^2\sim N(\z_i'\bbe,\tau^2), \\
&S_i^2|\si_i^2\sim \Ga\left(\frac{n_i-1}{2},\frac{n_i-1}{2\si_i^2}\right), \ \ \ \si_i^2\sim IG\left(a_i,b_i\ga\exp(\w_i'\bta)\right)\\
&\pi(\bbe,\tau^2,\ga,\bta)\propto 1,
\end{split}
\end{equation}
Then, the joint posterior distribution (\ref{pos}) is changed as
\begin{equation}\label{pos2}
\begin{split}
\pi(\theta_1,\ldots,\theta_m,&\si_1^2,\ldots,\si_m^2,\bbe,\tau^2,\ga,\bta|D)\propto (\tau^2)^{-m/2}\prod_{i=1}^{m}\ga^{a_i}\exp(a_i\w_i'\bta)(\si_i^2)^{-n_i/2-a_i-1}\\
&\times \exp\left\{-\frac{(X_i-\theta_i)^2+(n_i-1)S_i^2+2b_i\ga\exp(\w_i'\bta)}{2\si_i^2}-\frac{(\theta_i-\z_i'\bbe)^2}{2\tau^2}\right\}.
\end{split}
\end{equation}

We state our second main result, which provides a sufficient condition for the propriety of the posterior distribution given in (\ref{pos2}).
To this end, we define 
$$
t_k=\sgn\left(\sum_{i=1}^ma_iw_{ik}\right)\sgn\left(\sum_{i=1}^mn_iw_{ik}\right), \ \ \ k=1,\ldots,q,
$$
where $\sgn(x)$ for the real number $x$ denotes the sign of $x$.

\begin{thm}\label{thm:pos2} 
{\rm (a)}\ \ The marginal posterior density $\pi(\bbe,\tau^2,\ga,\bta|D)$ is proper if $m>p+2$, $n_i>1$, ${\rm rank}(\Z)=p$, and $t_k=1$ for $k=1,\ldots,q$.\\
{\rm (b)}\ \ The model parameters $\bbe,\tau^2, \ga$ and $\bta$ have finite posterior variances if $m>p+6$, $n_i>1$, ${\rm rank}(\Z)=p$, and $t_k=1$ for $k=1,\ldots,q$.
\end{thm}

The last new condition $t_k=1$ for $k=1,\ldots,q$ given in both (a) and (b) means that the two values $\sum_{i=1}^ma_iw_{ik}$ and $\sum_{i=1}^mn_iw_{ik}$ have the same signs for $k=1,\ldots,q$, while other conditions are the same as in Theorem \ref{thm:pos}.
Note that the simple sufficient condition for the last condition is $w_{ik}, \ i=1,\ldots,m$ have the same signs since $a_i$ and $n_i$ are positive.

To sample from the joint posterior distribution (\ref{pos2}), we can again use the Gibbs sampling method.
Note that the full conditional distributions of $\th_i$'s, $\bbe$ and $\tau^2$ are the same as (\ref{full}), and these of $\si_i^2$ and $\ga$ are obtained by replacing $b_i$ with $\exp(\w_i'\bta)$.
The full conditional distribution of $\bta$ is proportional to 
$$
\pi(\bta|\bsi^2,\ga,D)=\prod_{i=1}^m\exp(a_i\w_i'\bta)\exp\left\{-\frac{b_i\ga\exp(\w_i'\bta)}{\si_i^2}\right\},
$$
which is not a familiar form.
To sample from this full conditional distribution, we use the random-walk Metropolis-Hastings (MH) algorithm.
Let $\bta_0$ be the current value and we generate the proposal $\bta^{\ast}$ from $N_q(\bta_0,c \I_q)$ for specified $c>0$.
Then we accept the proposal $\bta^{\ast}$ with probability $\min\{1,p(\bta_0,\bta^{\ast})\}$, where
$$
p(\bta_0,\bta^{\ast})=\prod_{i=1}^m\exp\{a_i\w_i'(\bta^{\ast}-\bta_0)\}\exp\left(\frac{-b_i\gamma [\exp(\w_i'\bta^{\ast})-\exp(\w_i'\bta_0)]}{\si_i^2}\right).
$$

\section{Simulation studies}\label{sec:num}

In this section, we compare the accuracy of the hierarchical Bayes estimator based on the proposed full Bayesian model with the empirical Bayes estimator given by Maiti, et al. (2014) and the hierarchical model suggested in You and Chapman (2006) through simulation experiments.
We first generate observations for each small area from
$$
X_{ij}=\beta_0+\beta_1z_i+u_i+e_{ij}, \ \ \ j=1,\ldots,n_i, \ i=1,\ldots,m,
$$
where $u_i\sim N(0,\tau^2)$ and $e_{ij}\sim N(0,n_i\si_i^2)$.
Then the random effects model for the small area mean is 
$$
X_i=\beta_0+\beta_1z_i+u_i+e_i, \ \ \ \ i=1,\ldots,m,
$$
where $X_i=\overline{X}_i=n_i^{-1}\sum_{j=1}^{n_i}X_{ij}$ and $e_i=n_i^{-1}\sum_{j=1}^{n_i}e_{ij}$.
Therefore, $X_i|\theta_i\sim N(\theta_i,\si_i^2)$, where $\theta_i=\beta_0+\beta_1z_i+u_i$, that is $\theta_i\sim N(\beta_0+\beta_1z_i, )$, and $e_i\sim N(0,\si_i^2)$.
The parameter of interest is the mean $\theta_i$ in the $i$-th small area.
The direct estimator of $\si_i^2$ we used in simulation runs is
$$
S_i^2=\frac{1}{n_i(n_i-1)}\sum_{j=1}^{n_i}(X_{ij}-\overline{X}_i)^2,
$$
noting that $S_i^2|\si_i^2\sim \Ga((n_i-1)/2,(n_i-1)/2\si_i^2)$.
We generate covariate $z_i$ from the uniform distribution on $(2,8)$, and set the true parameter values $\beta_0=0.5, \beta_1=0.8$ and $\tau^2=1$.
We consider the case $m=30$ and $n_i=7$ for all areas.
For the true values of $\si_i^2$, we consider two cases: (i) $\si_i^2\sim IG(10,5\exp(0.3z_i))$ and (ii) $\si_i^2\sim U(0.5,5)$.

For simulated data, we apply four methods to get the estimator of the small area mean $\th_i$ and variance $\si_i^2$.
Two of four are the proposed Bayesian models (\ref{model}) and (\ref{var2}) referred as STK1 and STK2, respectively.
In applying these models, we put $a_i=2$ and $b_i=1/n_i$ as discussed in the end of Section \ref{sec:Bayes}, and we use $c=(0.2)^2$ in each MH step in STK2.
The third method is the hierarchical Bayesian method given by You and Chapman (2006) referred to as YC, where we assign the uniform prior for $\si_i^2$, namely $\pi(\si_i^2)\propto 1$.
For posterior sampling in YC method, we replace the full conditional for $\si_i^2$ in (\ref{full}) with
$$
\si_i^2|\bbe,\tau^2,\bsi^2_{(-i)},\bth,D\sim IG\left(\frac{n_i}{2},\frac12(X_i-\theta_i)^2+\frac12(n_i-1)S_i^2\right), \ \ i=1,\ldots,m, 
$$
and the propriety of the posterior distribution can be easily established from small modification of the proof of Theorem \ref{thm:pos}.
The fourth method is the empirical Bayes method given in Maiti, Ren and Sinha (2014) referred to as MRS.
In the three full Bayesian model, we calculate the estimators $\thh_i$ and $\sih_i^2$ as the mean of $5,000$ posterior samples after $1,000$ iteration.
For all four estimator, we calculate the mean squared errors and the absolute biases defined as $$
{\rm MSE}=\frac1{mR}\sum_{i=1}^m\sum_{r=1}^R(\thh_i^{(r)}-\th_i^{(r)})^2, \ \ \ \ {\rm Bias}=\frac1{mR}\sum_{i=1}^m\bigg|\sum_{r=1}^R(\thh_i^{(r)}-\th_i^{(r)})\bigg|,
$$
based on $R=2,000$ simulation runs, where $\thh_i^{(r)}$ and $\th_i^{(r)}$ are the estimated and true value in the $i$-th area in the $r$-th iteration.
Moreover, for the three Bayesian models STK1, STK2 and YC, we compute the credible intervals of $\th_i$ with probability $0.95$ and $0.99$, and calculated the coverage probability $(mR)^{-1}\sum_{i=1}^m\sum_{r=1}^RI(\th_i\in \widehat{{\rm CI}_{i(r)}})$, where $\widehat{{\rm CI}_{i(r)}}$ denotes the credible interval for $\th_i$ in the $r$-th run.
The simulation results are presented in Table \ref{tab:sim}.
For point estimation of $\th_i$, the MSEs of $\th_i$ in MRS is reasonable values, but the bias of MRS is larger compared to other three Bayesian models.
Among the full Bayesian model, it is observed that STK1 and STK2 attain minimum values of MSE in the case (ii) and (i), respectively.
The preference of YC is worst among the four models since YC does not consider the shrinkage estimation of $\si_i^2$ in spite of small sample sizes ($n_i=7$).
We also noted that the MSEs of $\si_i^2$ are largest in MRS in both cases, which may comes from instability of estimation of $\alpha$ and $\gamma$ in (\ref{maiti}).
Concerned with the Bayesian credible intervals, it is revealed that the suggested two methods STK1 and STK2 almost attain the nominal levels, but YC provides smaller coverage provabilities than the nominal levels.
This is clear that this phenomena comes from the instability of variance estimation in the YC method.
Therefore, the suggested procedure reasonably works in terms of MSE and bias of both $\th_i$ and $\si_i^2$, and can provide an accurate credible interval compared to the YC method.

\begin{table}[htb]
\caption{Simulation Result.
\label{tab:sim}
}
\begin{center}
\vspace{0.3cm}
\begin{tabular}{cccccccccccccc}
\toprule
&&& \multicolumn{2}{c}{Mean $(\th_i)$} & \multicolumn{2}{c}{Variance $(\si_i^2)$} & \multicolumn{2}{c}{CP} \\
&&& MSE & Bias & MSE & Bias & 95$\%$ & 99$\%$ \\
\midrule
(i)&STK1 && 1.120 & 0.036 & 2.325 & 0.411 & 95.6 & 99.3 \\
&STK2 && 1.102 & 0.035 & 2.087 & 0.272 & 95.3 & 99.2 \\
&YC && 1.275 &  0.038 & 3.894 & 0.120 & 93.2 & 97.6\\
&MRS && 1.149 & 0.410 & 4.442 &0.451 & --- & ---\\
\midrule
(ii)&STK1 && 1.043 & 0.040 & 1.144 & 0.041 & 95.2 & 99.2 \\
&STK2 && 1.053 & 0.041 & 1.845 & 0.278 & 95.5 & 99.4 \\
&YC && 1.185 & 0.044 & 2.630 & 0.099 & 93.0 & 97.9\\
&MRS && 1.001 & 0.273 & 2.849 & 0.320 & --- & ---\\
\bottomrule
\end{tabular}
\end{center}
\end{table}

\section{Real Data Analysis}\label{sec:emp}

\subsection{Survey data}

We apply the suggested procedures to the data in the Survey of Family Income and Expenditure (SFIE) in Japan. 
In this study, we use the data of the spending item `Education' (scaled by 1,000) in the survey in November 2011. 
The average spending at each capital city of 47 prefectures in Japan is denoted by $X_i$ for $i=1,\ldots,47$.
Although the average spendings in SFIE are reported every month, the sample sizes $n_i$'s are around 100 for most prefectures, and data of the item `Education' have high variability. 
On the other hand, we have data in the National Survey of Family Income and Expenditure (NSFIE) for 47 prefectures. 
Since NSFIE is based on much larger sample than SFIE, the average spendings in NSFIE are more reliable, but this survey has been implemented every five years.
In this study, we use the data of the item `Education' of NSFIE in 2009 as a covariate, which is denoted by $z_i$ for $i=1,\ldots,47$.
Then the two stage model for $X_i$ is described as
$$
X_i|\th_i,\si_i^2\sim N(\th_i,\si_i^2), \ \ \ \th_i|\be_0,\be_1,\tau^2\sim N(\beta_0+\beta_1z_i,\tau^2), \ \ \ i=1,\ldots,47.
$$
As the direct estimates of $\si_i^2$, we calculate $S_i^2$ from the data of the spending `Education' at the same city every November in the past ten years.
Then the model for $S_i^2$ is given by
$$
S_i^2|\si_i^2\sim \Ga\left(\frac{n_i-1}{2},\frac{n_i-1}{2\si_i^2}\right), \ \ \ \ i=1,\ldots,47.
$$
and the priors for $\si_i^2$ are given by
$$
{\rm (STK1)} \ \si_i^2\sim IG(a_i,b_i\gamma), \ \ \  {\rm (STK2)} \ \si_i^2\sim IG(a_i,b_i\gamma\exp(\eta z_i)), \ \ \ {\rm (YC)} \ \pi(\si_i^2)\propto 1.
$$
Remember that the uniform prior for $\si_i^2$ in YC model leads to the non-shrinkage posterior estimator of $\si_i^2$, while the proper prior for $\si_i^2$ in STK1 and STK2 leads to the shrinkage estimator of $\si_i^2$ toward the prior mean.

It is easy to confirm that the sufficient conditions in Theorems \ref{thm:pos} and \ref{thm:pos2} are satisfied in this case since the covariate $z_i$ is positive for all areas.
Now, we apply the three models to the survey data with 
$$
a_i=2\quad {\rm and}\quad b_i=1/n_i\quad \text{ in STK1 and STK2}.
$$
Moreover, to investigate sensitivity of the choices of $a_i$ and $b_i$, we consider the following two additional choices:
\begin{equation}\label{ch}
{\rm (s1)}\ \ \ a_i=3, \ b_i=1/n_i, \ \ \ \ \ \ \ \ {\rm (s2)} \ \ \ a_i=2, \ b_i=1,
\end{equation}
where the prior mean of $\si_i^2$ is $\ga/(2n_i)$ and $\ga$ in (s1) and (s2), respectively.
We use $c=1$ for MH step in STK2.
We first calculate the point estimates of model parameters as the means of $95,000$ posterior samples by Gibbs sampling after $5,000$ iteration.
The results are given in Table \ref{tab:survey}.
The estimated values of $\be_0, \be_1$ and $\tau^2$ are similar for all models.
For model comparison of these models, we calculated the Deviance Information Criterion (DIC) of Spiegelhalter, Best, Carlin and van der Linde (2002) given by ${\rm DIC}=2\overline{D(\phi)}-D(\overline{\phi})$, where $\bphi$ is the unknown model parameters, $D(\phi)$ is $(-2)$ times log-marginal likelihood function, and $\overline{D(\phi)}$ and $\overline{\phi}$ denote that posterior means of $D(\phi)$ and $\phi$, respectively.
Note that $\phi=\{\bbe,\tau^2,\gamma\}$ for STK1, $\phi=\{\bbe,\tau^2,\gamma,\bta\}$ for STK2, and 
$\phi=\{\bbe,\tau^2,\si_1^2,\ldots,\si_m^2\}$
for YC.
The resulting values of DIC and $\overline{D(\phi)}$ are reported in Table \ref{tab:survey}, and it is observed that YC is the most suitable model for this data set in terms of DIC.
This may come from the fact that the sample size $n_i$ in each area is around $100$.
Thus the direct estimates of sampling variances are relatively accurate in this case, so that it does not require shrinkage estimation for variances.
Comparing STK1, STK1-(s1) and STK1-(s2), $\gamma$ seems sensitive to the choice of $a_i$ and $b_i$, since the prior means are different for each choice, but the recommended choice attains the smaller value of DIC. 
The same thing can be observed in STK2, STK2-(s1) and STK2-(s2).
However, the posterior mean of $\th_i$ and $\si_i^2$ are nearly the same among the three choices.

\begin{figure}[htb]
\begin{center}
\includegraphics[width=8cm]{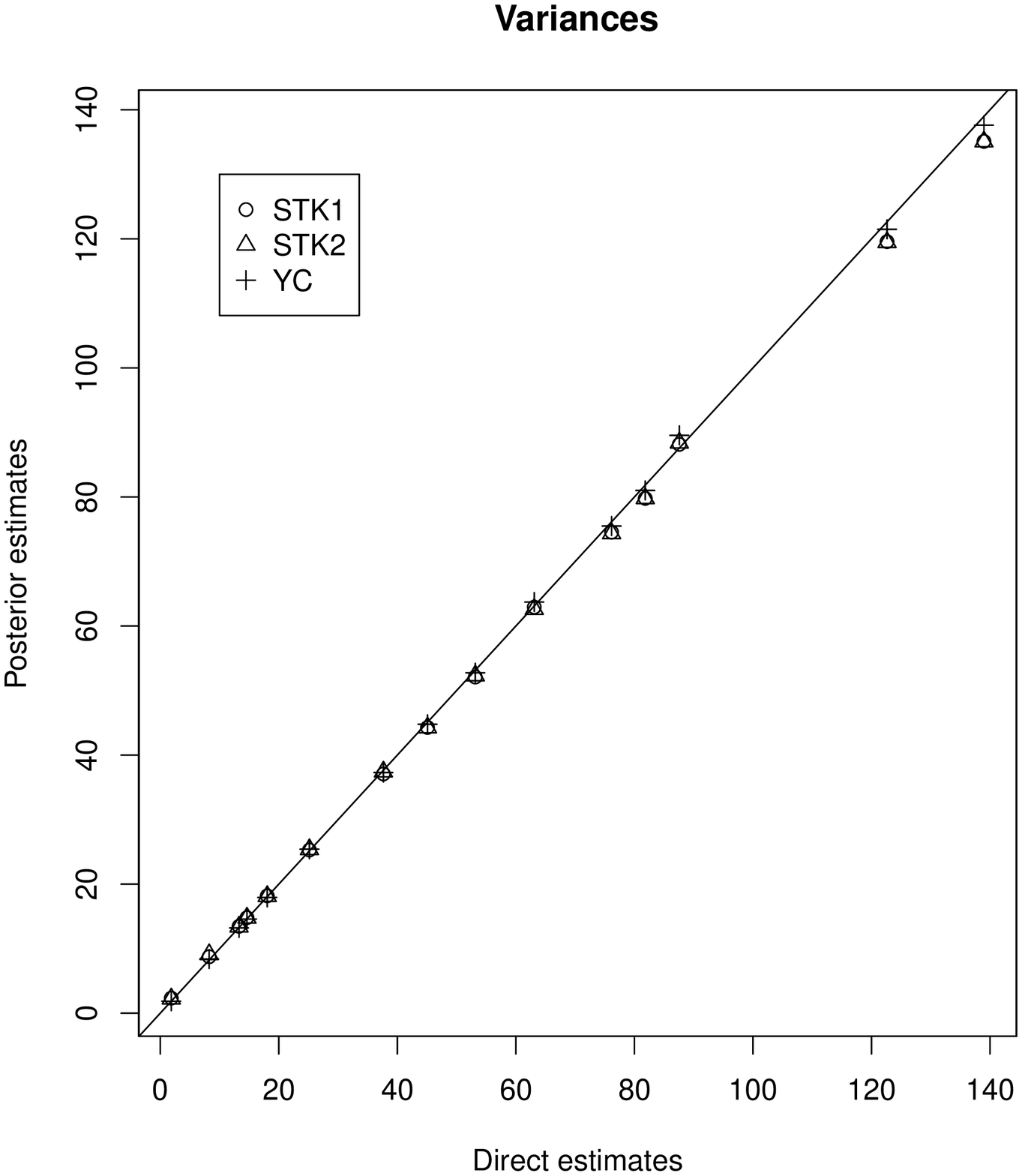}\ \ \ \ 
\includegraphics[width=8cm]{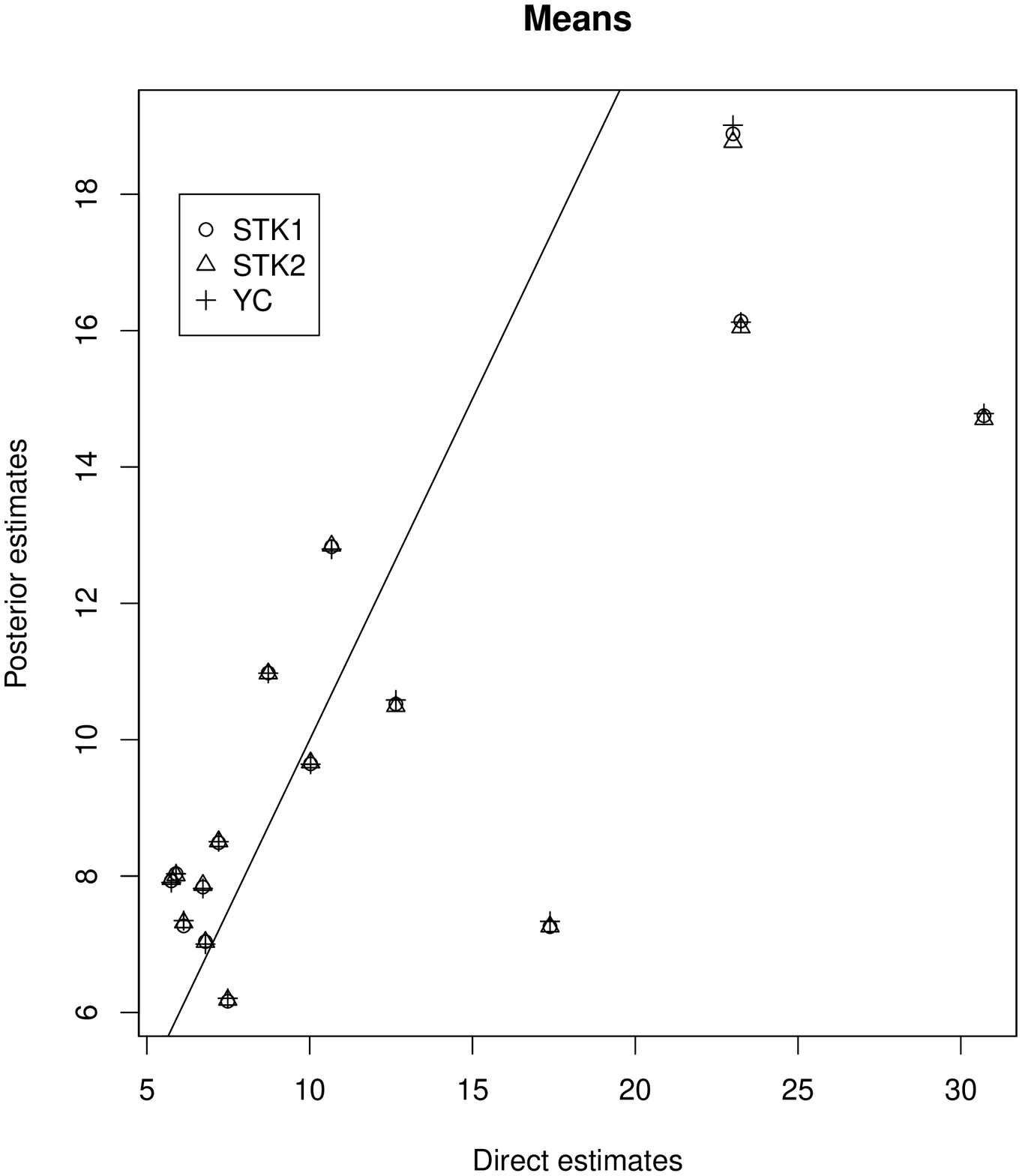}
\end{center}
\caption{Scatter Plots of Direct and Posterior Estimates of $\si_i^2$'s (Left) and $\th_i$'s (Right) for Selected 15 Areas in Survey Data.
\label{fig:survey}
}
\end{figure}

In the closing of this study, we compute the posterior estimates of $\si_i^2$'s and $\th_i^2$'s obtained from three models, STK1, STK2 and YC.
In Figure \ref{fig:survey}, we provide the scatter plots of direct and posterior estimates of $\si_i^2$'s and $\th_i$'s for selected 15 areas.
From the left panel of Figure \ref{fig:survey}, the posterior estimates of $\si_i^2$ are almost the same for each model in the area with small direct estimates.
On the other hand, in areas with large direct estimates of $\si_i^2$, the posterior estimates in YC and those of STK1 or STK2 are different since STK1 and STK2 produce shrinkage estimators for $\si_i^2$, but the difference is still small.
For the scatter plot for $\th_i$ given in the right panel of Figure \ref{fig:survey}, it is observed that the resulting posterior estimates from three models are similar.
Thus, the suggested procedures STK1 and STK2 provide almost the same estimates of $\th_i$, parameter of interest, as the YC method while the DIC values of STK1 and STK2 are larger than YC.
That is, both STK1 and STK2 work as well as YC in the case that there are no need to shrink direct estimates of variances.

\begin{table}[htb]
\caption{
Posterior Points Estimates and Standard Errors (Parenthesis) of Model Parameters, and DICs in Survey Data.
\label{tab:survey}
}
\begin{center}
\vspace{0.3cm}
\begin{tabular}{cccccccccccccc}
\toprule
& &  $\beta_0$ & $\beta_1$ & $\tau^2$ & $\gamma$ & $\eta$ & &  DIC  \\
 \midrule
STK1 & & 0.893 & 0.696 & 10.5 & $2.42\times 10^3$ &  --- && 700.4\\
          & & (2.74)&(0.206)&(5.15)& ($2.57\times 10^2$) &--- && &\\ 
STK1-(s1) & & 0.864 & 0.698 & 10.5 & $3.71\times 10^3$ &  --- && 717.4 \\
          & & (2.76)&(0.207)&(5.15)& ($3.29\times 10^2$) &--- && &\\
STK1-(s2) & & 0.918 & 0.694 & 10.4 & 23.6 &  --- && 705.0 \\
          & & (2.77)&(0.207)&(5.12)& (2.52) &--- && &\\
STK2 & & 0.868 & 0.697 & 10.4 & $1.22\times 10^3$& $5.47\times 10^{-2}$  && 700.3 \\
          &  &(2.78)&(0.209)& (5.17) &($3.08\times10^2$) & ($1.74\times 10^{-2}$) && &\\ 
STK2-(s1) & & 0.878 & 0.697 & 10.4 & $2.69\times 10^3$& $2.22\times 10^{-2}$  && 745.3  \\
          &  &(2.77)&(0.208)& (5.18) &($4.07\times10^2$) & ($1.08\times 10^{-2}$) && &\\ 
STK2-(s2) & & 0.831 & 0.700 & 10.6 & 30.6& $-1.68\times 10^{-2}$  && 6651.8  \\
          &  &(2.77)&(0.208)& (5.19) &(10.7) & ($2.63\times 10^{-2}$) && &\\ 
YC      & & 0.913 & 0.698 & 11.0& --- & --- && {\bf 558.3} \\
          & & (2.74)&(0.206) & (5.25) &--- & ---&&&\\
\bottomrule
\end{tabular}
\end{center}
\end{table}

\subsection{Corn data}
We next illustrate our methods based on the widely studied example which was first analyzed by Battese, Harter and Fuller (1988).
The dataset is on corn and soybean productions in 12 Iowa counties, and we here focus on corn data.
Since the sample size of the original data is ranging from 1 to 5, we cannot use the proposed model which requires $n_i>1$ for the posterior propriety as given in Theorem \ref{thm:pos}.
Thus, we use the modified data given in the table 6 in Dass, et al. (2012).
The dataset consists of $m=8$ areas with sample sizes in each area ranging from 3 to 5, and the survey data of corn ($X_i$) and the satellite data of both corn ($z_{1i}$) and soybeans ($z_{2i}$) as the covariates are observed in each area, where $X_i,z_{1i},z_{2i}$ are scaled by $100$.
Note that the sample sizes $n_i$ in each area is much smaller than that in the previous study.
Similarly to the previous study, we apply the three models STK1, STK2 with $a_i=2$ and $b_i=1/n_i$ and YC.
The two stage model for $X_i$ is given by
$$
X_i|\th_i,\si_i^2\sim N(\th_i,\si_i^2), \ \ \ \th_i|\be_0,\be_1,\be_2,\tau^2\sim N(\beta_0+\beta_1z_{1i}+\beta_2z_{2i},\tau^2), \ \ \ i=1,\ldots,8.
$$
For a covariate for variance modeling in STK2, we use only $z_{1i}$, namely $\si^2\sim IG(a_i,b_i\gamma\exp(\eta z_{1i}))$, since the DIC values of other models with use of only $z_{2i}$ and both $z_{1i}$ and $z_{2i}$ are larger than this model.
Since the covariate $z_{1i}$ is positive for all areas, the sufficient conditions in Theorem \ref{thm:pos} and \ref{thm:pos2} are satisfied in this case.
We use $c=(0.2)^2$ in each MH step in STK2.
We again consider two additional choices of $a_i$ and $b_i$ in (\ref{ch}).
Then, based on 95,000 posterior samples after 5,000 iteration, we calculate the point estimates of model parameters as the posterior sample means and we provide the resulting values in Table \ref{tab:corn} as well as DIC values.
The posterior estimates of regression coefficients $\be_0, \be_1$ and $\be_2$ are similar for all models, but $\ga$ and $\eta$ are different depending on the choices of $a_i$ and $b_i$.
It is also revealed that STK2 is the most preferable model for this data set from DIC values.
Among the three choices of $a_i$ and $b_i$, the recommended choice seems the best in terms of DIC, but the posterior mean of $\th_i$ and $\si_i^2$ are almost the same among the three choices.
In this case, it is interesting to point out that both STK1 and STK2 are more preferable than YC in terms of DIC values.
This is because the accuracy of the direct estimates of variances with small sample sizes (from $3$ to $5$) is suspicious and the shrinkage estimation for $\si_i^2$ is needed in this case.

\begin{figure}[htb]
\begin{center}
\includegraphics[width=8cm]{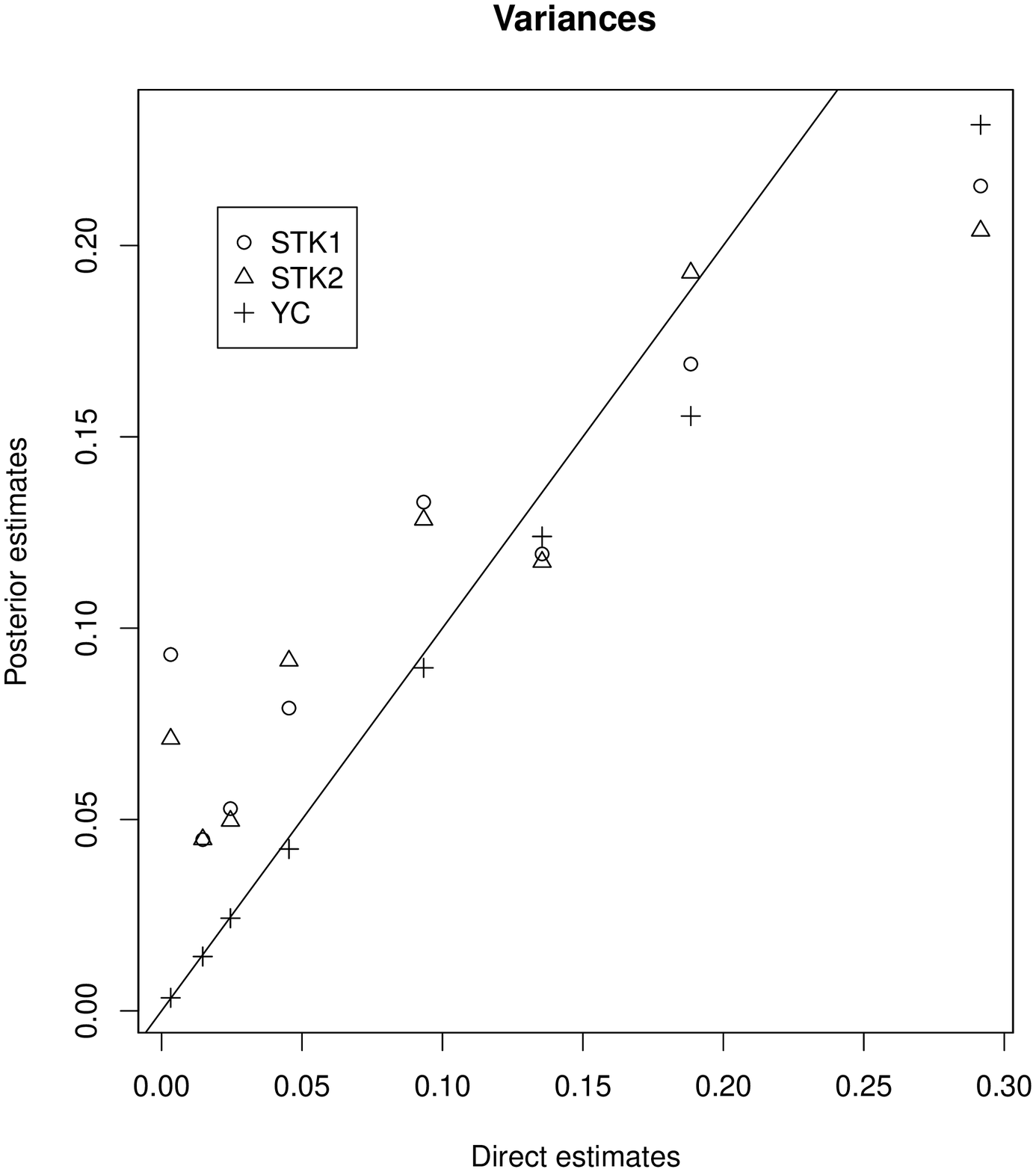}\ \ \ \ 
\includegraphics[width=8cm]{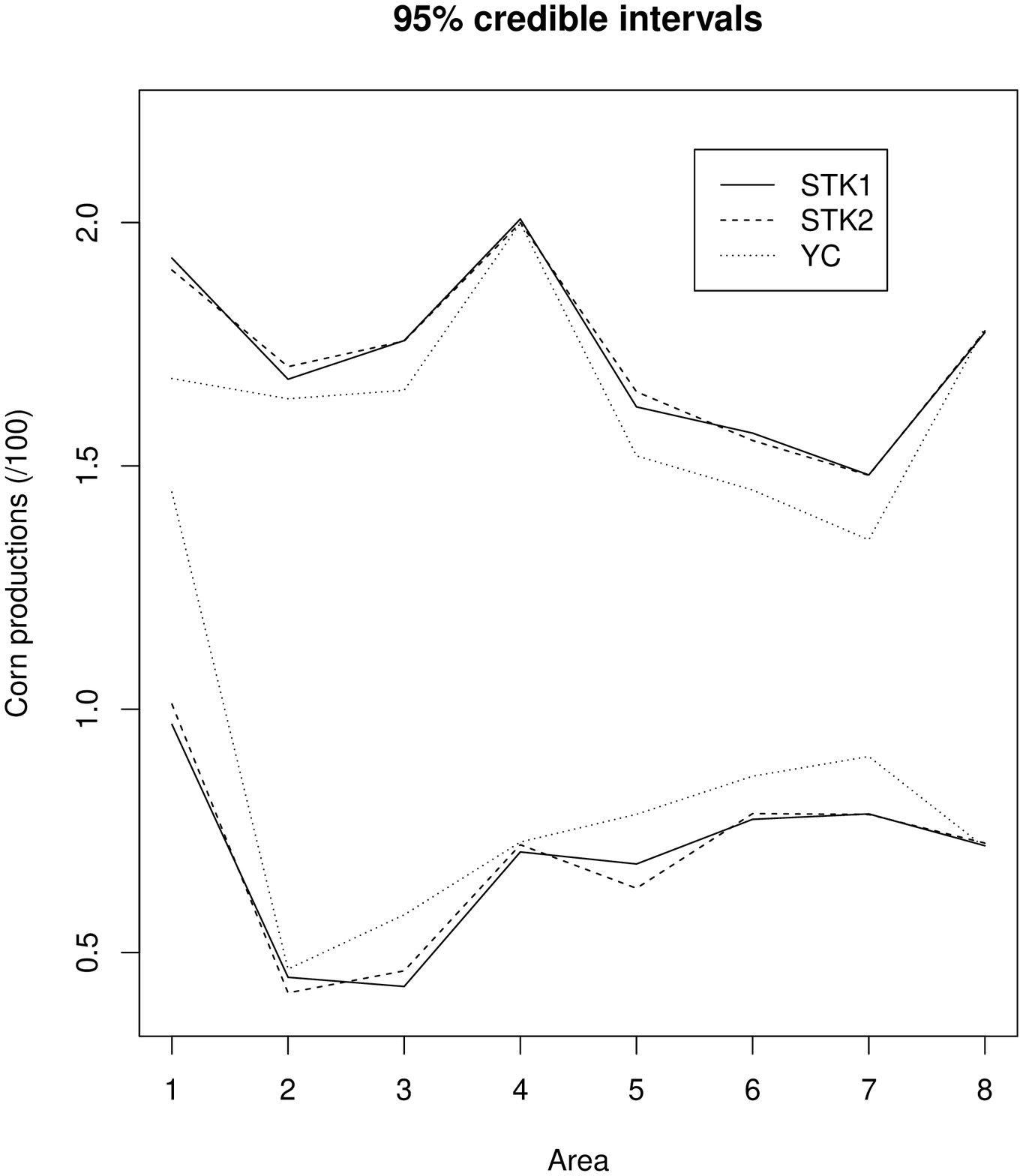}
\end{center}
\caption{Scatter Plots of Direct and Posterior Estimates of $\si_i^2$'s (Left) and $95\%$ Credible Intervals of $\th_i$'s (Right) in Corn Data.
\label{fig:corn}
}
\end{figure}

In the left panel of Figure \ref{fig:corn}, we show the scatter plots of direct and posterior estimates of $\si_i^2$ obtained from three models, STK1, STK2, and YC.
The result shows that the posterior estimates of $\si_i^2$ of YC (using uniform prior on $\si_i^2$) are considerably different from those of STK1 or STK2, while STK1 and STK2 produce the similar posterior estimated values.
It is also observed that the posterior estimator of $\si_i^2$ of STK1 and STK2 shrink the direct estimator of $\si_i^2$ toward some prior mean, but  that of YC does not.
In the right panel of Figure \ref{fig:corn}, we show the $95\%$ credible intervals for $\th_i$ from each model.
It is clear that STK1 and STK2 produce similar credible intervals and YC produces shorter credible intervals than two methods since the length of credible intervals are affected by the posterior estimates of $\si_i^2$.
In particular, the credible interval of YC in area 1 is much shorter than that of STK1 and STK2, but the interval of YC is not reliable because of instability of variance estimation in the YC method.
Then we may misinterpret the accuracy of the resulting estimator of $\th_i$ when we use YC in this case.
This phenomena is consistent to the simulation results in Table \ref{tab:sim}, where the credible interval in YC has smaller coverage probability than the true nominal level.
Thus the shrinking variances is the crucial strategy when $n_i$ is small like this data set.

\begin{table}[htb]
\caption{
Posterior Points Estimates and Standard Errors (Parenthesis) of Model Parameters, and DICs in Corn Data.
\label{tab:corn}
}
\begin{center}
\vspace{0.3cm}
\begin{tabular}{cccccccccccccc}
\toprule
& &  $\beta_0$ & $\beta_1$ & $\beta_2$ & $\tau^2$ & $\gamma$ & $\eta$ & &  DIC \\
 \midrule
STK1 & &  -1.59& 0.679 & 0.379 & 0.278 & 0.559 & --- && -14.45 \\
          &  & (9.47)&(1.97)&(1.88)&(1.25)& (0.252)&--- &&&\\ 
STK1-(s1)  &&  -1.42& 0.643 & 0.347 & 0.279 & 0.884 & --- && -14.23 \\
          &  & (9.68)&(2.02)&(1.89)&(1.69)&(0.367) &--- &&&\\ 
STK1-(s2) &&-1.73  & 0.726 & 0.385 & 0.341 & 0.144 & --- && -11.76\\
          &  & (9.67)&(2.03)&(1.90)&(2.63)&(0.0655) &--- &&&\\ 
STK2 & &  -1.76& 0.720 & 0.402 & 0.367 & 8.67 & -0.939 && {\bf -20.39} \\
          &  & (11.0)&(2.38)&(2.03)&(7.21)&(4.77) &(0.154) &&&\\
STK2-(s1) & &  -1.57& 0.686 & 0.358 & 0.256 & 14.5 & -0.961 && -10.02 \\
          &  & (9.06)&(1.90)&(1.77)&(0.821)&(7.05) &(0.118) &&&\\    
STK2-(s2) & &  -1.74& 0.729 & 0.384 & 0.283 & 7.31 & -1.27 && -3.10\\
          &  & (9.53)&(1.99)&(1.88)&(1.20)&(5.56) &(0.299) &&&\\  
YC     & &  -1.805& 0.754 & 0.375 & 0.303 & --- & --- && -7.33 \\
          &  & (9.57)&(1.99)&(1.88)&(1.11)&--- &--- &&&\\ 
\bottomrule
\end{tabular}
\end{center}
\end{table}

\section{Concluding remarks}\label{sec:conc}
In this paper, we have proposed the Bayesian small area models shrinking both means and variances.
As the empirical Bayes approach, Maiti, et al. (2014) proposed estimating the model parameters from the marginal likelihood function, but the marginal likelihood function is not obtained in a closed form, so that we need to rely on the EM algorithm including numerical integral evaluation in each iteration step.
On the other hand, the proposed Bayesian method does not suffer from the numerical complexity since all the full conditional posterior distributions are familiar forms as described in Section \ref{sec:B1}, and we can easily sample from the posterior distributions using Gibbs sampling.
Moreover, we have also suggested to use the covariate information in variance modeling as described in Section \ref{sec:var}.
In the Bayesian analogy, You and Chapman (2006) also proposed the Bayesian model with use of estimated sampling variances, but they used (almost) uniform prior for $\si_i^2$ and their model cannot produce the shrinkage estimator of $\si_i^2$.
Compared to You and Chapman (2006), we have used proper prior for $\si_i^2$ with unknown scale parameter to produce the shrinkage estimator of $\si_i^2$.
All Bayesian models suggested in this paper are objective since we use the uniform priors for unknown model parameters.
The validity of posterior inferences is guaranteed by the propriety of the posterior distributions and finite variances of the model parameters under mild sufficient conditions given in Theorems \ref{thm:pos} and \ref{thm:pos2}.

\ \\
{\bf Acknowledgement}

\medskip
We would like to thank the editor, the associate editor and two reviewers for many valuable comments and helpful suggestions which led to an improved version of this paper.
The first author was supported in part by Grant-in-Aid for Scientific Research (10076) from Japan Society for the Promotion of Science (JSPS).
Research of the third author was supported in part by Grant-in-Aid for Scientific Research  (15H01943 and 26330036) from Japan Society for the Promotion of Science.

\vspace{1cm}
\begin{center}
{\bf Appendix}
\end{center}

\vspace{0.5cm}
\noindent
{\bf Proof of Theorem \ref{thm:pos}.}\ \ \ \ We first prove part (a). 
Let $\Re_{+}=\{x\in\Re|\ x>0\}$ be the set of positive numbers.
In what follows, capital $C$, with and without suffix, means a generic constant.
It is sufficient to prove that 
$$
\int_{\Re^m\times \Re_{+}^2}\pi(\bbe,\tau^2,\ga|D)d\bbe d\tau^2 d\ga<\infty.
$$
Let $\bth=(\theta_1,\ldots,\theta_m)'$, $\bsi^2=(\si_1^2,\ldots,\si_m^2)'$.
Then we need to prove that 
$$
\int_{\Re^m\times \R_{+}^m\times \R^p\times \Re_{+}^2}\pi(\theta_1,\ldots,\theta_m,\si_1^2,\ldots,\si_m^2,\bbe,\tau^2,\ga|D)d\bth d\bsi^2 d\bbe d\tau^2d\ga<\infty,
$$
where 
\begin{equation*}
\begin{split}
\pi(\theta_1,&\ldots,\theta_m,\si_1^2,\ldots,\si_m^2,\bbe,\tau^2,\ga|D)\\
&\propto (\tau^2)^{-m/2}\prod_{i=1}^{m}\ga^{a_i}(\si_i^2)^{-n_i/2-a_i-1}\exp\left(-\frac{(X_i-\theta_i)^2+(n_i-1)S_i^2+2b_i\ga}{2\si_i^2}-\frac{(\theta_i-\z_i'\bbe)^2}{2\tau^2}\right).
\end{split}
\end{equation*}
From expression (\ref{pos}), we first integrate with respect to $\si_1^2,\ldots,\si_m^2$ to get
\begin{align*}
\pi(\bth,\bbe,\tau^2,\ga|D)\propto (\tau^2)^{-m/2}\exp\left(-\frac{(\bth-\Z'\bbe)'(\bth-\Z'\bbe)}{2\tau^2}\right)\prod_{i=1}^m\ga^{a_i}\psi_i(\theta_i-X_i,\ga)^{-(n_i/2+a_i)},
\end{align*}
where $\psi_i(\theta_i-X_i,\ga)=(X_i-\theta_i)^2+(n_i-1)S_i^2+2b_i\ga$.
Noting that 
$$
\int_{\Re^p}\exp\left(-\frac{(\bth-\Z\bbe)'(\bth-\Z'\bbe)}{2\tau^2}\right)d\bbe=(\tau^2)^{p/2}|\Z'\Z|^{-1/2}\exp\left\{-\frac1{2\tau^2}\bth'(\I_m-\Z(\Z'\Z)^{-1}\Z')\bth\right\},
$$
we obtain
\begin{equation}\label{A:tau}
\pi(\bth,\tau^2,\ga|D)
\propto (\tau^2)^{-(m-p-2)/2-1}\exp\left\{-\frac1{2\tau^2}\bth' \A \bth\right\}\prod_{i=1}^m\ga^{a_i}\psi_i(\theta_i-X_i,\ga)^{-(n_i/2+a_i)},
\end{equation}
for $\A=\I_m-\Z(\Z'\Z)^{-1}\Z'$.
When $m-p-2>0$ i.e. $m>p+2$, we can integrate (\ref{A:tau}) with respect to $\tau^2$ to get
\begin{equation}\label{A:theta}
\pi(\bth,\ga|D)\propto\left(\bth'\A\bth\right)^{-(m-p-2)/2}\prod_{i=1}^m\ga^{a_i}\psi_i(\theta_i-X_i,\ga)^{-(n_i/2+a_i)}.
\end{equation}
Making the transformation $\bmu=(\mu_1, \ldots, \mu_m)'=\bth-\X$, we have
\begin{equation}\label{A:pp1}
\pi(\bmu,\ga|D)
\propto\left\{(\bmu+\X)'\A(\bmu+\X)\right\}^{-(m-p-2)/2}\prod_{i=1}^m\ga^{a_i}\psi_i(\mu_i,\ga)^{-(n_i/2+a_i)}.
\end{equation}
Since $\A$ is an idempotent matrix with ${\rm rank}(\A)=m-p$, there exists a $(m-p)\times m$ matrix $\H_1$ such that $\A=\H_1'\H_1$ and $\H_1\H_1'=\I_{m-p}$.
Then, $(\bmu+\X)'\A(\bmu+\X)=\bmu'\H_1'\H_1\bmu + 2 \X'\H_1'\H_1\bmu + \X'\H_1'\H_1\X$.
Let $\P$ be a $(m-p)\times (m-p)$ orthogonal matrix such that $\P'=(\P_1', \P_2')$ and $\X'\H_1'\P_2'=\0'$ where $\P_1$ is a $1\times (m-p)$ vector.
Since $\I_{m-p}=\P_1'\P_1+\P_2'\P_2$, it is observed that
\begin{align*}
(\bmu+\X)'\A(\bmu+\X)=&\bmu'\H_1'(\P_1'\P_1+\P_2'\P_2)\H_1\bmu + 2 \X'\H_1'\P_1'\P_1\H_1\bmu + \X'\H_1'(\P_1'\P_1+\P_2'\P_2)\H_1\X
\\
=&
(\bmu +\X)'\H_1'\P_1'\P_1\H_1(\bmu+\X) + \bmu'\H_1'\P_2'\P_2\H_1\bmu + \X'\H_1'\P_2'\P_2\H_1\X
\\
\geq & \X'\H_1'\P_2'\P_2\H_1\X,
\end{align*}
which is used to evaluate $\int \pi(\bmu,\ga|D)d\bth d\ga$ from above as
\begin{align}
\int_{\Re^m\times \Re_{+}}\pi(\bmu,\ga|D)d\bmu d\ga
<&C \int_{\Re^m\times \Re_{+}}\left(\X'\H_1'\P_2'\P_2\H_1\X\right)^{-(m-p-2)/2}\prod_{i=1}^m\ga^{a_i}\psi_i(\mu_i,\ga)^{-(n_i/2+a_i)} d\bmu d\ga
\nonumber\\
=&C' \int_0^\infty \prod_{i=1}^m \Big\{ \int_{-\infty}^\infty \ga^{a_i}\psi_i(\mu_i,\ga)^{-(n_i/2+a_i)} d\mu_i \Big\} d\ga.\label{eval}
\end{align}
Making the transformation $u_i=\mu_i/\sqrt{(n_i-1)S_i^2+2b_i\ga}$ gives 
$$
\int_{-\infty}^\infty \ga^{a_i}\psi_i(\mu_i,\ga)^{-(n_i/2+a_i)} d\mu_i 
= {\ga^{a_i} \over \{ (n_i-1)S_i^2+2b_i\ga \}^{(n_i-1)/2+a_i}} \int_{-\infty}^\infty {1\over (1+u_i^2)^{n_i/2+a_i}}d u_i.
$$
Note that $\int_{-\infty}^\infty (1+u_i^2)^{-B}d u_i = 2 \int_0^\infty (1+u_i^2)^{-B}d u_i \leq 2\int_0^1 (1+u_i^2)^{-B}d u_i + 2\int_1^\infty u_i^{-2B}d u_i$, which is finite if $2B>1$.
Thus, 
$$
\int_{-\infty}^\infty {1\over (1+u_i^2)^{n_i/2+a_i}}d u_i < \infty
$$
since $n_i+2a_i>0$.
Noting that $\ga^{a_i}/\{ (n_i-1)S_i^2+2b_i\ga \}^{(n_i-1)/2+a_i} \leq (2b_i)^{-(n_i-1)/2}/\{ (n_i-1)S_i^2+2b_i\ga \}^{(n_i-1)/2}$, we can see that
$$
\int_{\Re^m\times \Re_{+}}\pi(\bmu,\ga|D)d\bmu d\ga
\leq C \int_{0}^{\infty}\left\{(n_{\ast}-1)S_{\ast}^2+2b_{\ast}\ga\right\}^{-(N-m)/2}d\ga<\infty,
$$
where $N=\sum_{i=1}^mn_i$ and $n_{\ast},S_{\ast}^2,b_{\ast}$ are the minimum values of $\{n_i\},\{S_i^2\},\{b_i\}$, respectively.
This is finite for $N-m>2$.
Thus the proof for part (a) is complete.

\medskip
For part (b), we show $E(\bbe\bbe'|D), E((\tau^2)^2|D)$ and $E(\ga^2|D)$ are finite.
For $E((\tau^2)^2|D)$, we evaluate it in the same manner as in Part (a).
Note that
\begin{equation*}
\begin{split}
(\tau^2)^2\pi(\bth,\tau^2,\ga|D)\propto (\tau^2)^{-(m-p-6)/2-1}\exp\left(-\frac1{2\tau^2}\bth'\A\bth\right)\prod_{i=1}^m\ga^{a_i}\psi_i(\theta_i-X_i,\ga)^{-(n_i/2+a_i)},
\end{split}
\end{equation*}
so that it follows, when $m-p-6>0$, namely $m>p+6$, that
$$
E((\tau^2)^2|D)<C \int_{\Re^{m}\times\Re_{+}}\prod_{i=1}^m\ga^{a_i}\psi_i(\theta_i-X_i,\ga)^{-(n_i/2+a_i)}d\bth d\ga<\infty.
$$
For evaluating $E(\bbe\bbe'|D)$, note that
\begin{align*}
\int_{\Re^p}&\bbe\bbe'\exp\left(-\frac{(\bth-\Z\bbe)'(\bth-\Z'\bbe)}{2\tau^2}\right)d\bbe\\
&=(\tau^2)^{p/2}|\Z'\Z|^{-1/2}\exp\left(-\frac1{2\tau^2}\bth'\A\bth\right)(\Z'\Z)^{-1}\left\{\tau^2\I_m+\Z'\bth\bth'\Z(\Z'\Z)^{-1}\right\}.
\end{align*}
Integrating out it with respect to $\tau^2$, we have
\begin{align*}
E(\bbe\bbe'|D)
\propto& \int_{\Re^m\times\Re_{+}}\left(\bth'\A\bth\right)^{-(m-p-4)/2}\prod_{i=1}^m\ga^{a_i}\psi_i(\theta_i-X_i,\ga)^{-(n_i/2+a_i)} d\bth d\ga (\Z'\Z)^{-1}\\
&+\int_{\Re^m\times\Re_{+}}{ (\Z'\Z)^{-1}\Z'\bth\bth'\Z(\Z'\Z)^{-1} \over (\bth'\A\bth)^{-(m-p-2)/2}}\prod_{i=1}^m{\ga^{a_i}\over \psi_i(\theta_i-X_i,\ga)^{n_i/2+a_i}} d\bth d\ga.
\end{align*}
The first term can be verified to be finite, since we can use the same arguments as in (\ref{A:theta}), (\ref{A:pp1}) and (\ref{eval}).
For the second term, we make the transformation $\bmu=\bth-\X$ to rewrite it as
$$
\int_{\Re^m\times\Re_{+}}{ (\Z'\Z)^{-1}\Z'(\bmu+\X)(\bmu+\X)'\Z(\Z'\Z)^{-1} \over \{(\bmu+\X)'\A(\bmu+\X)\}^{-(m-p-2)/2}}\prod_{i=1}^m{\ga^{a_i}\over \psi_i(\mu_i,\ga)^{n_i/2+a_i}} d\bmu d\ga,
$$
so that it is sufficient to show that for $j=1, \ldots, m$,
\begin{equation}
\int_{\Re^m\times\Re_{+}}{ \mu_j^2 \over \{(\bmu+\X)'\A(\bmu+\X)\}^{-(m-p-2)/2}}\prod_{i=1}^m{\ga^{a_i}\over \psi_i(\mu_i,\ga)^{n_i/2+a_i}} d\bmu d\ga <\infty.
\label{evalbe}
\end{equation}
By the same arguments as (\ref{eval}), the inequality (\ref{evalbe}) is satisfied if
\begin{align}
\int_0^\infty \Big\{\int_{-\infty}^\infty {\ga^{a_j}\mu_j^2 \over \psi_i(\mu_j,\ga)^{n_j/2+a_j}} d\mu_j \Big\} \prod_{i\not= j} \Big\{ \int_{-\infty}^\infty {\ga^{a_i}\over \psi_i(\mu_i,\ga)^{n_i/2+a_i-1}} d\mu_i \Big\} d\ga<\infty.\label{evalbe1}
\end{align}
Making the transformation $u_j=\mu_j/\sqrt{(n_j-1)S_j^2+2b_j\ga}$ gives 
$$
\int_{-\infty}^\infty {\ga^{a_j}\mu_j^2 \over \psi_i(\mu_j,\ga)^{n_j/2+a_j}} d\mu_j 
= {\ga^{a_j} \over \{ (n_j-1)S_i^2+2b_j\ga \}^{(n_j-3)/2+a_j}} \int_{-\infty}^\infty {u_j^2\over (1+u_j^2)^{n_j/2+a_j}}d u_j,
$$
which is finite since $n_j>1$.
Hence, the inequality (\ref{evalbe1}) is satisfied if
$$
\int_{0}^{\infty}\left\{(n_{\ast}-1)S_{\ast}^2+2b_{\ast}\ga\right\}^{-K/2}d\ga<\infty,
$$
where $K=n_j-3+\sum_{i\neq j}(n_i-1)=N-m-2$.
This establishes that $E(\bbe\bbe'|D)<\infty$ for $N>m+4$.
Finally, for $E(\ga^2|D)$, it follows that for $N>m+6$,
$$
E(\ga^2|D)<C\int_{0}^{\infty}\ga^2\left\{\frac12(n_{\ast}-1)S_{\ast}^2+b_{\ast}\ga\right\}^{-(N-m)/2}d\ga<\infty,
$$
which completes the proof for (b).

\vspace{1cm}
\noindent
{\bf Proof of Theorem \ref{thm:pos2}.} \ \ \ \ 
We first prove part (a).
From (\ref{eval}) given in the proof of Theorem \ref{thm:pos}, it is sufficient to show that 
\begin{equation}\label{pf2}
\int_{\Re_{+}\times \Re^q}\prod_{i=1}^m\left(\ga\exp(\w_i'\bta)\right)^{a_i}\Big\{\frac12(n_i-1)S_i^2+b_i\ga\exp(\w_i'\bta)\Big\}^{-(n_i/2+a_i)}d\ga d\bta<\infty,
\end{equation}
under the condition that $t_k=1$ for $k=1,\ldots,q$.
Since $(n_i-1)S_i^2$ and $\ga\exp(\w_i'\bta)$ are positive, the left side in (\ref{pf2}) is evaluated from the upper by
\begin{equation}\label{pf3}
\int_{\Re_{+}\times \Re^q}\ga^{A}\prod_{k=1}^q\exp(\eta_k)^{B_{1k}}\Big\{C_{\ast}+b_{\ast}\ga^{A+N/2+m}\prod_{k=1}^q\exp(\eta_k)^{B_{1k}+B_{2k}}\Big\}^{-1}d\ga d\bta,
\end{equation}
where $A=\sum_{i=1}^ma_i$, $B_{1k}=\sum_{i=1}^ma_iw_{ik}$, $B_{2k}=2^{-1}\sum_{i=1}^mn_iw_{ik}$, $b_{\ast}=\prod_{i=1	}^mb_i^{-(n_i/2+a_i)}$, and $C_{\ast}=2^{-(A+N/2+m)}\prod_{i=1}^m\{(n_i-1)S_i^2\}^{-(n_i/2+a_i)}$.
Thus we need to show that (\ref{pf3}) is finite.
Without loss of generality, we consider the case of $B_{1k}>0$ and $B_{2k}>0$ for $k=1,\ldots,q$, since  the case that $B_{1k}<0$ and $B_{2k}<0$ for some $k$ reduces to $B_{1k}>0$ and $B_{2k}>0$ by changing the variable $\eta_k$ as $-\eta_k$.
From the positivity of $B_{1k}$'s, there exists $\la>0$ such that $B_{1k}>1/\la>0$ for $k=1,\ldots,q$, and we change the variables as $\phi_k=\exp(\eta_k/\la)$ in (\ref{pf3}) to get $\int_{\Re_{+}^{q+1}} f(\ga,\bphi) d\ga d\bphi$, where $\bphi=(\phi_1, \ldots, \phi_q)$ and
$$
f(\ga,\bphi)=\la^q\ga^{A}\prod_{k=1}^q\phi_k^{\la B_{1k}-1}\Big(C_{\ast}+b_{\ast}\ga^{A+N/2+m}\prod_{k=1}^q\phi_k^{\la B_{1k}+\la B_{2k}}\Big)^{-1}.
$$
We decompose the integral $\int_{\Re_{+}^{q+1}} f(\ga,\bphi) d\ga d\bphi$ into the $2^{q+1}$ domains $\ga\leq 1$ or $\ga\geq1$, and $\phi_k\leq 1$ or $\phi_k\geq 1$ for $k=1,\ldots,q$.
Then it is sufficient to show that 
\begin{equation}\label{integrate}
\int_{0}^1  \int_{(0,1]^r\times  [1,\infty)^{q-r}}  f(\ga,\bphi)d\bphi d\ga<\infty, \ \ \ \ \ \int_{1}^{\infty} \int_{(0,1]^r\times  [1,\infty)^{q-r}} f(\ga,\bphi)d\bphi d\ga<\infty,
\end{equation}
for fixed $r=0,\ldots,q$.
For evaluating the former in (\ref{integrate}), we define $g(\ga,\phi_1,\ldots,\phi_r)=\int_{[1,\infty)^{q-r}}f(\ga,\bphi)d\bphi$.
We note that $g(\ga,\phi_1,\ldots,\phi_r)$ is $0$ when at least one among $\ga,\phi_1,\ldots,\phi_r$ is 0, and $g(\ga,\phi_1,\ldots,\phi_r)<\infty$ for other values since 
\begin{align*}
g(\ga,\phi_1,\ldots,\phi_r)&=
\la^q\ga^{A}\prod_{k=1}^r\phi_{k}^{\la B_{1k}-1}\int_{[1,\infty)^{q-r}}\prod_{k=r+1}^q\phi_k^{\la B_{1k}-1}\Big(C_{\ast}+D_{\ast}\prod_{k=r+1}^q\phi_k^{\la B_{1k}+\la B_{2k}}\Big)^{-1}d\phi_{r+1}\ldots d\phi_{q}\\
&\leq 
\la^q\ga^{A}\prod_{k=1}^r\phi_{k}^{\la B_{1k}-1}D_{\ast}^{-1}\prod_{k=r+1}^q\int_{1}^{\infty}\phi_k^{-\la B_{2k}-1}d\phi_k<\infty,
\end{align*}
for $0<\ga,\phi_{1},\ldots,\phi_{r}\leq 1$, where $D_{\ast}=b_{\ast}\ga^{A+N/2}\prod_{k=1}^r\phi_{k}^{\la B_{1k}+\la B_{2k}}$.
Therefore, $g(\ga,\phi_1,\ldots,\phi_r)$ is bounded over $[0,1]^r$, so that the former integral in (\ref{integrate}) is finite.
For the latter case of (\ref{integrate}), we can similarly show that the integral is finite since $N/2>1$, which completes the proof for part (a).

\medskip
For part (b), we first note that it can be proved of finiteness of the posterior variances of other parameters using the similar argument given in the proof of part (a) in Theorem \ref{thm:pos2}. 
Hence, we show $E[\bta_k^2|D], k=1,\ldots,q$ are finite.
To this end, it is sufficient to prove that 
\begin{equation*}
\int_{\Re_{+}\times \Re^q}\ga^{A}\eta_k^2\prod_{\ell=1}^q\exp(\eta_\ell)^{B_{1\ell}}\Big\{C_{\ast}+b_{\ast}\ga^{A+N/2}\prod_{k=1}^q\exp(\eta_k)^{B_{1\ell}+B_{2\ell}}\Big\}^{-1}d\ga d\bta<\infty,
\end{equation*}
for $k=1,\ldots,q$.
Under the condition that $B_{1k}>0$ and $B_{2k}>0$ for $k=1,\ldots,q$, there exists $\la>0$ such that $B_{1k}>3/\la$ and $B_{2k}>3/\la$, and we change the variables as $\phi_k=\exp(\eta_k/\la)$ in the left side to get $\int_{\Re_{+}^{q+1}} f_{k}(\ga,\bphi) d\ga d\bphi$, where
$$
 f_{k}(\ga,\bphi)=\la^3\ga^{A}\prod_{\ell=1}^q(\log\phi_k)^2\phi_\ell^{\la B_{1\ell}-1}\Big\{C_{\ast}+b_{\ast}\ga^{A+N/2}\prod_{\ell=1}^q\phi_\ell^{\la B_{1\ell}+\la B_{2\ell}}\Big\}^{-1}.
$$
We again decompose the $2^{q+1}$ domains $\ga\leq 1$ or $\ga\geq1$, and $\phi_k\leq 1$ or $\phi_k\geq 1$ for $k=1,\ldots,q$.
Since $\la B_{1\ell}-1>2$, $(\log\phi_k)^2\phi_k^{\la B_{1k}-1}$ is bounded over $0<\phi_k\leq 1$.
On the other hand, it is noted that $\int_1^\infty (\log \phi_k)^2 \phi_k^{\la B_{1\k}-1}/(C+D\phi_k^{\la B_{1k}+\la B_{2k}}) d\phi_k = \int_{0}^{\infty}u^2\exp(\la B_{1k}u)/(C+D\exp\{(\la B_{1k}+\la B_{2k})u\})du<\infty$ under $B_{2k}>0$.
Therefore, similar evaluation shows that the integral $\int_{\Re_{+}^{q+1}} f_{k}(\ga,\bphi) d\ga d\bphi$ is finite, whereby we complete the proof for part (b).


\end{document}